\documentclass[12pt]{article}
\usepackage[numbers,sort&compress]{natbib}
\usepackage{hyperref}
\usepackage{geometry}
\usepackage{graphicx}
\usepackage{authblk}
\usepackage{caption}
\usepackage{float}
\usepackage{xcolor}
\usepackage{booktabs}
\usepackage{tabularx}
\usepackage{array}
\usepackage{enumitem}
\usepackage{microtype}
\usepackage{amsmath}
\usepackage{amssymb}
\usepackage{tikz}
\usetikzlibrary{arrows.meta,positioning,calc,shapes.geometric,fit,backgrounds}
\begin{document}

\title{Preparing Students for AI-Powered Materials Discovery:\\A Workflow-Aligned Framework for AI Literacy, Equity, and Scientific Judgment}

\author[1]{Dongming Mei\thanks{Corresponding author: \texttt{dongming.mei@usd.edu}}}
\author[2]{Katherine Moore}
\author[3]{Ben Sayler}

\affil[1]{Department of Physics, University of South Dakota\\ 414 E. Clark St.\\ Vermillion, SD 57069, USA}
\affil[2]{Massachusetts Institute of Technology\\ 77 Massachusetts Avenue\\ Cambridge, MA 02139, USA}
\affil[3]{Black Hills State University\\ 1200 University Street\\ Spearfish, SD 57799\textendash 9005, USA}

\setlength{\affilsep}{0.6em}
\renewcommand\Authfont{\normalsize}
\renewcommand\Affilfont{\small}
\date{\today}
\maketitle

\begin{abstract}
Artificial intelligence (AI) is reshaping education, scientific training, and materials
discovery. In materials science, AI models increasingly support property prediction,
experiment prioritization, and hypothesis generation; however, the limiting factor is
no longer only algorithmic capability but also whether students and educators can
use AI with domain-specific scientific judgment. This workshop-informed white paper 
and curriculum-oriented position article argues that AI education for AI-powered materials 
discovery must move beyond tool access and surface-level interaction with generative AI systems toward a workflow-aligned 
model of AI literacy. We connect AI literacy to materials-informatics competencies: data 
provenance, domain-specific featurization, model validation, uncertainty quantification, 
physics-informed reasoning, reproducibility, and experimental feedback. We also 
emphasize outcome-oriented equity: institutions should evaluate not only access, 
participation, and engagement, but also whether AI-enabled instruction produces 
comparable learning gains, transfer of learning, confidence calibration—the alignment 
with students’ confidence and the quality or correctness of their work—persistence, 
and research readiness across student subgroups. The paper synthesizes relevant 
evidence, identifies risks for learners such as cognitive off-loading and cognitive surrender, 
and provides a dual-track curriculum model and implementation recommendations such
as curriculum guides and an assessment plan for 
courses, bootcamps, workshops, and program-level reform. The central goal is to prepare 
students to become better scientists—not merely more efficient users of AI tools.

\end{abstract}

\vspace{1em}

\section{Introduction}

Artificial intelligence (AI) has moved rapidly from research laboratories into everyday products, services, and educational environments. \emph{Generative AI} refers to systems that produce new text, images, code, or other content from prompts based on patterns learned from training data. \emph{Machine learning} (ML) refers to algorithms that learn patterns from data to make predictions, classifications, recommendations, or decisions. Together, these technologies now influence how students write, code, search for information, solve problems, and receive feedback. They also influence how scientists screen materials, prioritize experiments, interpret complex datasets, and generate hypotheses. The motivation for this paper was strengthened by the \emph{Workshop for AI-Powered Materials Discovery at Great Plains}, held at the University of South Dakota from June~22--25, 2025, with funding from the National Science Foundation~\cite{AIMaterialsWorkshop2025}. The workshop brought together more than 200 participants and highlighted a clear regional and national need for AI literacy, workforce development, responsible data practices, and domain-specific training in AI-enabled materials discovery.

This paper is a workshop-informed white paper and curriculum-oriented position article. It does not present new empirical data. Instead, it synthesizes research and practice across AI education, cognitive science, intelligent tutoring systems, and materials informatics, and translates that synthesis into a practical framework for curriculum design and evaluation. We argue that AI education for materials discovery must move beyond generic tool use. Students should not merely learn to ask a chatbot for an answer or run a model in an interactive computational notebook.\footnote{Interactive computational notebooks combine executable code, data analysis, visualization, explanatory text, and reproducible workflow documentation in a single electronic environment (e.g., Jupyter Notebook).} Rather, they should learn to explain where data came from, why a representation is physically meaningful, how a model was validated, what uncertainty remains, whether a prediction is experimentally plausible, and how AI use affects learning across student subgroups. The primary audiences for this framework are faculty designing AI-enabled materials courses, program leaders developing data-centric science curricula, workshop organizers building short-form training modules, and education researchers evaluating AI-supported learning. The framework is intended to support both incremental adoption within existing courses and broader program-level reform.

Discussions at the workshop emphasized that AI is not simply a new computational tool. To use AI responsibly, students and researchers must understand basic algorithms, the limitations and biases embedded in data, the role of domain knowledge in model interpretation, and the ethical questions raised by automation. AI-literacy frameworks similarly emphasize that students should engage with AI-enhanced methods while maintaining critical thinking, scientific judgment, and human accountability~\cite{Fengchun2024,Kassorla2024,Mills2024}. These needs are especially urgent in materials science, where AI tools increasingly influence how researchers select candidate materials, prioritize experiments, interpret high-dimensional data, and connect predictions to physical mechanisms.

In materials science, the convergence of AI and materials research is often described as \emph{materials informatics} (MI): an interdisciplinary field that uses data science, statistics, and ML to represent materials, predict properties, and guide experiments. MI is accelerating the discovery of functional compounds and process routes, from battery electrolytes and electrocatalysts to high-performance alloys. Yet the dominant bottleneck is increasingly \emph{educational}. MI requires students and researchers to blend rigorous physical-science reasoning with modern data and machine-learning practices. This is not a ``generic coding'' problem. The field demands domain-specific representations of chemistry and structure, careful validation under small-data constraints, and physics-informed methods that respect thermodynamic, structural, and mechanistic limits~\cite{Oweida2020,Wang2020,DeCost2020}. At the same time, cultural and cognitive barriers---including skepticism of probabilistic models among experimentally trained experts---can slow adoption even when methods are useful~\cite{Boyce2023}.

Adoption of AI in education is also growing rapidly. Survey evidence indicates that many U.S. college students began using generative AI soon after the release of tools such as ChatGPT, and many students report using AI for information gathering and brainstorming rather than only for shortcuts or misconduct~\cite{cengage2025}. Faculty attitudes are more mixed: instructors see potential benefits for personalization and feedback, but they also report concerns about academic integrity, bias, accuracy, and lack of training~\cite{cengage2025}. These survey findings are important evidence of scale and urgency, but they do not by themselves demonstrate learning impact. The educational challenge is therefore not simply whether students and faculty will use AI, but whether institutions can guide AI use toward deeper understanding, better transfer of learning, stronger scientific reasoning, and more equitable outcomes.

\begin{figure}[H]
\centering
\begin{tikzpicture}[
    font=\small,
    phase/.style={draw=#1!65!black, rounded corners, thick, align=center,
        minimum width=3.25cm, minimum height=1.05cm, fill=#1!9},
    center/.style={draw=blue!70!black, rounded corners, very thick, align=center,
        minimum width=3.8cm, minimum height=1.25cm, fill=blue!10},
    arrow/.style={-{Latex[length=2.8mm]}, very thick, blue!70!black},
    support/.style={-{Latex[length=2.0mm]}, thick, dashed, gray!70}
]
\node[center] (c) at (0,0) {\textbf{AI-powered}\\\textbf{materials discovery}};
\node[phase=cyan]   (data)  at (0, 2.7) {\textbf{1. Data}\\experiments, simulations,\\literature};
\node[phase=orange] (repr)  at (4.5,0) {\textbf{2. Representation}\\descriptors, featurization,\\provenance};
\node[phase=purple] (model) at (0,-2.7) {\textbf{3. Modeling}\\validation, uncertainty,\\prediction};
\node[phase=green]  (exp)   at (-4.5,0) {\textbf{4. Experiment}\\synthesis, testing,\\feedback};
\draw[arrow] (data.east) to[bend left=20] (repr.north);
\draw[arrow] (repr.south) to[bend left=20] (model.east);
\draw[arrow] (model.west) to[bend left=20] (exp.south);
\draw[arrow] (exp.north) to[bend left=20] (data.west);
\draw[support] (c) -- (data);
\draw[support] (c) -- (repr);
\draw[support] (c) -- (model);
\draw[support] (c) -- (exp);
\end{tikzpicture}
\caption{Workflow-aligned competency map for AI-powered materials discovery. The closed loop connects data provenance and quality, representation and featurization, modeling and validation, uncertainty-aware decision making, and experimental synthesis/testing. The same loop structures the curriculum and assessment recommendations in this paper.}
\label{fig:aimd_loop}
\end{figure}
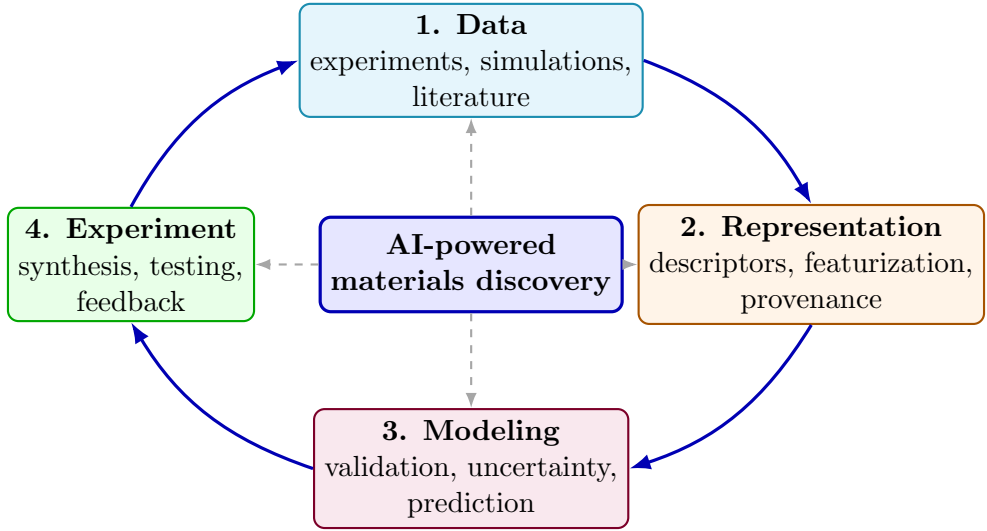

Figure~\ref{fig:aimd_loop} summarizes the closed-loop workflow that underpins AI-powered materials discovery. A student who can train a generic model but cannot judge whether a descriptor is physically meaningful, whether a data split leaks chemical-system information, whether uncertainty is calibrated, or whether a prediction violates known thermodynamic constraints is not yet prepared for AI-powered materials discovery. Conversely, a student who understands only the physical science but not data provenance, validation, and model uncertainty may be unable to evaluate AI-generated claims. This loop clarifies why the central bottleneck is educational rather than purely algorithmic.

The novelty of this framework lies in its integration of three elements that are often treated separately: AI literacy, materials-informatics competency, and outcome-oriented equity. Rather than proposing AI literacy as generic tool fluency, we align learning outcomes with the materials discovery workflow and require evidence that AI-supported instruction improves scientific reasoning and research readiness across student subgroups.

This paper makes four contributions beyond prior general AI-literacy discussions:
\begin{itemize}[leftmargin=*]
  \item \textbf{Materials-informatics competency framing:} We define what AI literacy means in the context of AI-powered materials discovery, emphasizing domain-specific representation, small-data realities, uncertainty quantification, physics-informed modeling, and experimental validation.
  \item \textbf{Workflow-aligned education model:} We map educational competencies to the end-to-end materials discovery loop and identify where human judgment, verification, physical constraints, and ethical governance enter the workflow.
  \item \textbf{Actionable curriculum artifacts:} We provide an example eight-week MI module sequence, assignments, and a compact rubric that can be deployed in existing courses, bootcamps, workshops, or graduate training programs.
  \item \textbf{Outcome-oriented equity framework:} We distinguish access equity from impact equity and recommend disaggregated evaluation of learning gains, transfer of learning, confidence calibration (the alignment between students' confidence and the correctness or quality of their work), persistence, and research readiness across student subgroups.
\end{itemize}

Together, these contributions motivate the curricular and governance recommendations developed below. The remainder of the paper reviews the current state of AI in education, outlines critical needs for AI literacy and MI training, summarizes benefits and risks, proposes implementation strategies, and provides reusable curriculum artifacts and evaluation tools.

\section{Current State of AI in Education and Key Definitions}

\subsection{The educational landscape}

The use of AI in classrooms has expanded dramatically. Survey evidence shows broad student adoption of generative language models such as ChatGPT, with students reporting use for brainstorming, information gathering, drafting, revision, and homework support~\cite{cengage2025}. The same reports indicate that many students believe they know more about AI than their instructors and want explicit guidance on AI skills in relevant courses~\cite{cengage2025}. Faculty concerns remain substantial, especially around academic integrity, bias, accuracy, and lack of training~\cite{cengage2025}. These data establish an important starting point: AI use is already widespread, but institutional guidance and pedagogical capacity remain uneven.

Beyond generative text models, AI is embedded in a range of educational technologies. \emph{Intelligent tutoring systems} (ITS) are AI-enabled learning environments that model a learner's knowledge state and adapt instruction to learner needs. Recommendation engines curate videos and readings, voice-assistant tutors support language learning, adaptive assessments adjust difficulty in real time, and learning analytics systems track student progress and flag potential disengagement. These tools show promise, but their adoption is uneven and often limited by infrastructure, cost, teacher preparation, institutional policy, and the availability of high-quality curricular content.

\subsection{Evidence streams and equity framing}

This paper draws on evidence streams that differ in purpose and strength, and we interpret them accordingly. \emph{Meta-analyses and systematic reviews} provide the strongest evidence for average effects and moderators across multiple studies. \emph{Experimental and quasi-experimental studies} offer stronger causal inference about specific interventions but may be context-dependent because effects vary by task, guardrails, classroom integration, learner expertise, and assessment design. \emph{Surveys and industry reports} describe adoption patterns, perceptions, and institutional readiness; they are valuable for understanding scale and urgency, but they should not be treated as proof of learning impact. \emph{Policy guidance documents} from agencies, states, and professional organizations provide normative frameworks for responsible use, governance, and competency design; they are essential for implementation but do not by themselves demonstrate effectiveness. Throughout this paper, we therefore use cautious language: survey evidence suggests adoption and concern, experimental evidence indicates causal effects under defined conditions, and meta-analytic evidence supports broader claims about average learning effects.

The current landscape also requires a sharper equity lens. \emph{Access equity} asks whether students, teachers, and institutions can obtain devices, broadband, software, datasets, and support. \emph{Impact equity} asks whether AI-enabled learning produces comparable gains in conceptual understanding, transfer of learning, self-regulation, applied competence, and persistence across subgroups and institutional contexts. Both are necessary. Access without learning gains can reproduce inequality in a more technologically sophisticated form, while learning gains observed only in average outcomes can obscure subgroup harms.

\section{Critical Needs for AI Literacy and Integration}

Students need more than superficial exposure to AI---for example, learning to ``use'' a generative tool without understanding what it is doing, what data it was trained on, and how its outputs can fail. Multiple organizations and scholars have therefore articulated AI literacy frameworks: structured sets of competencies, learning outcomes, and assessment practices that balance \emph{technical knowledge} with \emph{practical wisdom}~\cite{Fengchun2024,Kassorla2024,Mills2024,DEC2025,OECD2025}. Technical knowledge includes how algorithms, data pipelines, model training, and evaluation work. Practical wisdom includes when and how to deploy AI appropriately, ethically, and with human accountability.

Across these frameworks, AI literacy is not a single skill but a progression of competencies: (i) conceptual understanding of learning systems, including classification, regression, overfitting, uncertainty, and bias--variance tradeoffs; (ii) data competence, including data quality, provenance, documentation, and responsible sharing; (iii) evaluation competence, including validity, robustness, and task-appropriate metrics; and (iv) socio-technical competence, including fairness, privacy, transparency, human oversight, and accountability. UNESCO's AI competency framework for students emphasizes technical foundations, social--ethical awareness, and responsible innovation~\cite{Fengchun2024}. Kassorla et al. connect algorithmic principles to critical pedagogy in higher education~\cite{Kassorla2024}. Mills et al. frame AI literacy as an evolving capability for both learners and educators~\cite{Mills2024}. Policy-oriented organizations, including the Digital Education Council and the OECD, provide guidance for translating literacy principles into curriculum, assessment, and governance structures~\cite{DEC2025,OECD2025}. At the K--12 level, state guidance is proliferating, but implementation remains uneven~\cite{AIForEducation2025}.

\subsection{A dual-track model for AI education}

Building on these frameworks, we recommend a dual-track educational approach: stand-alone AI foundations combined with AI practice embedded inside disciplinary courses. Foundational AI courses should teach core concepts such as data preparation, algorithm design, model selection, validation, uncertainty, bias detection, reproducibility, and responsible use. These courses build transferable competencies so students can reason about model performance, identify failure modes, and communicate limitations.

\begin{figure}[H]
\centering
\begin{tikzpicture}[
  font=\small,
  >=Latex,
  boxA/.style={
    draw=blue!65!black,
    rounded corners,
    thick,
    align=center,
    fill=blue!7,
    text width=3.35cm,
    minimum height=1.35cm,
    inner sep=5pt
  },
  boxB/.style={
    draw=green!55!black,
    rounded corners,
    thick,
    align=center,
    fill=green!8,
    text width=3.35cm,
    minimum height=1.35cm,
    inner sep=5pt
  },
  mid/.style={
    draw=orange!70!black,
    rounded corners,
    very thick,
    align=center,
    fill=orange!12,
    text width=3.15cm,
    minimum height=1.35cm,
    inner sep=6pt
  },
  mainarrow/.style={
    -{Latex[length=2.6mm]},
    very thick,
    blue!65!black
  },
  feedback/.style={
    <->,
    very thick,
    gray!70
  },
  lab/.style={
    font=\footnotesize,
    fill=white,
    inner sep=2.5pt,
    text=black
  }
]

\node[mid]  (grad)  at (0,0) {
  \textbf{AI-literate}\\
  \textbf{graduates}
};

\node[boxA] (found) at (-5.0,0) {
  \textbf{Foundational AI}\\
  concepts, data,\\
  models, validation
};

\node[boxB] (disc)  at (5.0,0) {
  \textbf{Disciplinary}\\
  \textbf{integration}\\
  materials, science,\\
  society, ethics
};

\draw[mainarrow]
  (found.east) -- 
  node[lab, above=3pt] {core tools}
  (grad.west);

\draw[mainarrow]
  (disc.west) -- 
  node[lab, above=3pt] {authentic context}
  (grad.east);

\draw[feedback]
  (found.north) .. controls (-3.7,2.05) and (3.7,2.05) ..
  node[lab, above=4pt] {shared projects}
  (disc.north);

\draw[feedback]
  (disc.south) .. controls (3.7,-2.05) and (-3.7,-2.05) ..
  node[lab, below=4pt] {evidence from applications}
  (found.south);

\end{tikzpicture}

\caption{Dual-track approach to AI education. Foundational AI courses and disciplinary integration jointly produce AI-literate graduates; they also reinforce each other through shared projects, contextual feedback, and evidence from authentic applications.}
\label{fig:dualtrack}
\end{figure}
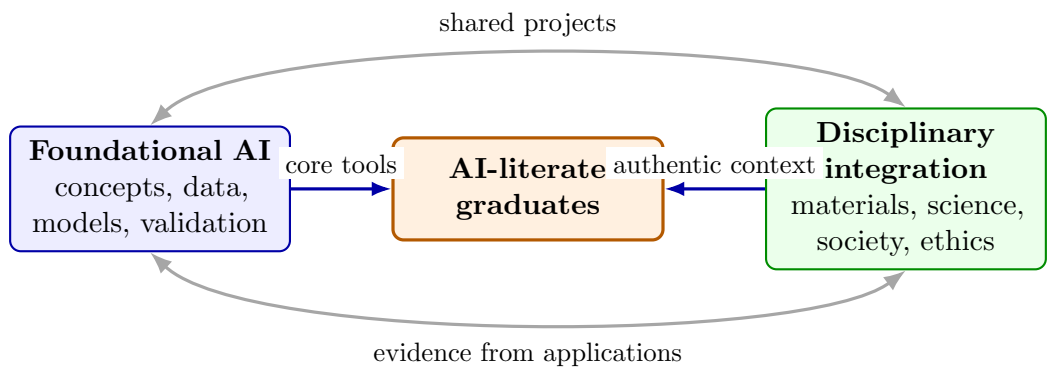

The second track embeds AI inside disciplinary learning. Students should encounter AI not only in a computing course but also in physics, chemistry, biology, engineering, humanities, social science, and education contexts. This allows them to see how assumptions, data quality, ground truth, metrics, uncertainty, and ethical concerns change by domain~\cite{CasalOtero2023,CrawfordWu2024,Southworth2023}. The two tracks should not operate as separate silos. Foundational AI courses provide conceptual and technical tools; disciplinary integration teaches students how those tools behave in real knowledge-making environments. Figure~\ref{fig:dualtrack} shows this mutually reinforcing structure.

Evidence from the education literature supports this structure. Casal-Otero et al.'s review of K--12 AI literacy highlights developmental alignment and project-based learning~\cite{CasalOtero2023}. Jackson discusses the need to close the ``new digital divide'' by bringing AI concepts into K--12 computer-science education rather than limiting them to well-resourced schools~\cite{Jackson2025}. More broadly, an integrated approach helps students learn not only \emph{how} to use AI tools, but also \emph{when} to rely on them, \emph{how} to verify outputs, and \emph{how} to maintain human responsibility in high-stakes decisions.

\subsection{Materials informatics as an educational bottleneck}

General AI literacy is necessary but insufficient for materials discovery. A student may understand classification, regression, or prompt engineering in general terms yet still fail to recognize when a materials dataset mixes incompatible measurement protocols, when a descriptor encodes spurious chemical-system information, or when a predicted property violates known physical constraints. Materials-informatics education therefore requires AI literacy to be embedded in the epistemic practices of materials science: measurement, representation, uncertainty, mechanism, and experimental validation.

Teaching AI for materials discovery is uniquely challenging because it must blend deep physical science knowledge with modern data-centric methods. In contrast to many mainstream ML applications, materials problems often involve heterogeneous data modalities---composition, structure, process history, microstructure, simulations, spectra, images, and measured properties---and comparatively limited, noisy experimental datasets. As a result, MI education must be deliberately domain-specific rather than a generic programming appendix to the traditional materials curriculum~\cite{Oweida2020}.

A central premise of this paper is that AI literacy for materials discovery should be evaluated by demonstrated scientific judgment, not merely by tool use. Students should be able to explain where data came from, why a representation is physically meaningful, how a model was validated, what uncertainty remains, and whether a prediction is experimentally plausible. Similarly, equity in AI education should be measured not only by access to software or participation in AI-enabled activities, but also by whether students from different backgrounds achieve comparable gains in conceptual understanding, transfer of learning, confidence calibration, and research readiness. This outcome-oriented view links AI literacy, MI competence, and educational equity into a single evaluation framework.

\subsubsection{Core curriculum competencies}

\textbf{Domain-specific featurization.} Featurization means translating physical, chemical, and structural knowledge into machine-readable variables. A central learning objective in MI is constructing descriptors that preserve relevant invariances, such as translation, rotation, atom indexing, and symmetry, while enabling reliable performance on materials not seen during training. Because descriptor choices are often user-dependent, students should compare composition-based, structure-based, graph-based, and learned representations and evaluate descriptor quality, sensitivity, and physical interpretability~\cite{Oweida2020}.

\textbf{The ``nuts and bolts'' of machine learning.} MI programs must cover data cleaning, preprocessing, algorithm selection, baseline construction, hyperparameter tuning, and robust validation. Students should be able to compare model classes, interpret performance metrics, and diagnose failure modes such as data leakage, overfitting, and distribution shift~\cite{Wang2020}. These skills matter because a model can appear accurate under a random split while failing when asked to generalize to new chemical systems or processing regimes.

\textbf{Physics-informed AI and scientific interpretation.} Physics-informed AI incorporates scientific constraints, such as conservation laws, symmetries, thermodynamic bounds, and mechanistic priors, into learning and inference. In MI education, students should learn to detect unphysical extrapolations, distinguish correlation from mechanism, and connect predictions to experimentally testable hypotheses~\cite{DeCost2020}. The goal is not only prediction but also scientific understanding.

\textbf{Uncertainty-aware decision making.} Materials datasets are often sparse and expensive to expand. Students therefore need to understand uncertainty quantification, calibration, and active learning. They should be able to ask whether a model is uncertain because a candidate material lies outside the training distribution, whether uncertainty estimates are calibrated against observed errors, and how uncertainty should guide the next experiment or simulation~\cite{Lookman2019ActiveLearning}.

\subsubsection{Educational bottlenecks and systemic challenges}

\textbf{Curriculum gaps and faculty shortages.} Traditional materials science programs often lack systematic data science integration. A survey of materials programs reported limited availability of courses that explicitly address data handling and machine learning, with one driver being the shortage of cross-trained faculty who can teach both materials fundamentals and computational methods~\cite{Oweida2020}.

\textbf{The small-data reality.} Unlike commercial AI systems that can train on massive, uniform datasets, materials datasets are frequently sparse, noisy, expensive to generate, and inconsistent across laboratories. MI education must train students to reason under uncertainty, manage missing metadata, document provenance, and use careful cross-validation, uncertainty quantification, and active learning to maximize learning from limited data~\cite{Oweida2020,Wang2020}.

\textbf{Disciplinary skepticism toward probabilistic models.} Materials research culture rightly values mechanistic explanation and gold-standard characterization. Resistance to statistical models is often justified when interpretability is weak, data quality is unclear, or predictions conflict with known physics. Effective MI education should therefore teach model trust as an evidence-based practice: students should learn when to trust, when to verify, when to reject, and when to redesign a model~\cite{Boyce2023}.

\subsubsection{Agile pedagogical strategies that work now}

Because multi-year degree redesign is slow, the MI community has adopted agile strategies to bridge skills gaps. Interactive notebooks allow students to reproduce and modify end-to-end ML workflows without weeks of software setup~\cite{Wang2020}. Intensive workshops and bootcamps, including short courses at Materials Research Society (MRS) meetings, provide rapid upskilling for students and active researchers who did not receive formal MI training~\cite{MannodiKanakkithodi2023}. These formats are especially useful for regional collaborations because they can be shared across institutions, adapted to local datasets, and evaluated with common rubrics.

Professional development for educators is another critical pillar. Surveys indicate that many teachers already use generative AI for lesson planning or administration, yet many feel unprepared to integrate AI into pedagogy~\cite{NEA2025,Langreo2024}. The National Education Association calls for sustained professional learning communities, and Tan's scoping review argues that teacher development in the generative AI era requires iterative, practice-based training rather than one-off workshops~\cite{NEA2025,Tan2025}. Programs such as AI Book Clubs---structured professional learning communities in which educators read, discuss, test, and adapt AI resources together---provide one model for sustained professional learning~\cite{Lee2022,Zhang2023,Zhang2024}.

\section{Benefits of AI in Education}

When thoughtfully designed, AI can enhance learning by augmenting rather than replacing human educators. A mature example is the class of intelligent tutoring systems. ITS aim to approximate some benefits of one-on-one tutoring by modeling a learner's evolving knowledge state and delivering targeted practice and feedback. Meta-analytic evidence indicates that ITS can improve learning outcomes, particularly when systems provide immediate, actionable feedback and when instruction is aligned with clear learning objectives~\cite{VanLehn2011,KulikFletcher2016}. Beyond correctness feedback, well-designed tutors often include step-level guidance, worked-example support, and metacognitive prompts that encourage learners to explain their reasoning rather than simply obtain an answer~\cite{ChiWylie2014}.

The strongest educational case for AI is not replacement of instructors but orchestration of learning. In learning-analytics contexts, AI systems can help teachers identify where students struggle, decide when whole-class instruction or small-group intervention is needed, and revise curricula based on evidence from student interaction data. Learning analytics can collect interaction traces---for example, time on task, hint use, error patterns, and concept mastery estimates---that help educators identify bottlenecks not visible from summative grades alone. AI-powered analytics can also flag struggling students and recommend interventions, while adaptive assessments can select items that better estimate mastery and reduce measurement error~\cite{VanLehn2011,KulikFletcher2016}. At the program level, aggregate data can support evidence-based curriculum revision and help identify concepts where particular student populations encounter barriers.

Generative AI and adaptive learning tools can also support students directly when used as scaffolds---temporary supports that help students perform a task while still requiring their own reasoning---rather than as substitutes for learning. AI tools can provide targeted practice, formative feedback, adaptive pacing, accessibility supports such as speech-to-text, text-to-speech, translation, captioning, and assistive writing support, and prompts that ask students to articulate assumptions, compare alternative explanations, critique model outputs, or revise their reasoning in response to feedback. In this role, generative AI is most valuable when it preserves student agency and requires students to explain, verify, and extend their own reasoning.

In higher education, AI can support authentic research experiences by accelerating data analysis, simulation, and hypothesis generation. In materials-focused curricula, this can enable undergraduates and early graduate students to participate in modern workflows such as screening candidate materials, interpreting structure--property relationships, and prioritizing experiments within closed-loop discovery cycles~\cite{Butler2018,Lookman2019ActiveLearning}. These opportunities are especially valuable when students are asked to document data provenance, evaluate model limitations, and explain the physical plausibility of AI-generated recommendations.

AI integration can also help reduce geographic and socio-economic barriers to learning. In rural or remote communities where private tutoring, advanced coursework, or specialized instructors may be limited, AI-powered tutoring systems can provide structured support and additional practice beyond the constraints of classroom time~\cite{VanLehn2011,KulikFletcher2016}. However, the equity benefits of AI are not automatic and should not be inferred from access alone. A tutoring system that increases log-in rates but produces smaller learning gains for rural students, English learners, first-generation students, students with disabilities, or students from under-resourced schools would not be equitable in an instructional sense. Institutions therefore need to evaluate whether AI support improves conceptual understanding, transfer of learning, self-regulation, persistence, and research readiness across subgroups, while also addressing device access, broadband connectivity, digital literacy, privacy, and teacher preparation.

AI can also reduce some forms of educator workload. Routine tasks such as drafting feedback templates, organizing practice sets, generating formative summaries, and tracking student progress can be partially automated. If used responsibly, this can free instructors to focus more time on mentoring, relationship building, conceptual clarification, and higher-value instructional interactions---the human-centered aspects of teaching that are difficult to automate and central to student success.

\begin{figure}[H]
\centering
\begin{tikzpicture}[
  font=\small,
  col/.style={draw, rounded corners, thick, align=left, minimum width=6.2cm, minimum height=4.7cm, inner sep=7pt},
  good/.style={col, draw=green!45!black, fill=green!5},
  risk/.style={col, draw=red!55!black, fill=red!4}
]
\node[good] (b) {\textbf{Potential benefits}\\[2mm]
$\checkmark$ targeted practice and feedback\\
$\checkmark$ adaptive pacing and assessment\\
$\checkmark$ accessibility supports\\
$\checkmark$ authentic research workflows\\
$\checkmark$ reduced routine workload};
\node[risk, right=1.1cm of b] (r) {\textbf{Risks requiring guardrails}\\[2mm]
$\triangleright$ cognitive off-loading and surrender\\
$\triangleright$ weaker transfer of learning\\
$\triangleright$ technostress and social isolation\\
$\triangleright$ breach of privacy, bias, and misinformation\\
$\triangleright$ unequal subgroup learning gains};
\end{tikzpicture}
\caption{Benefits and risks of AI-supported education. Educational value is maximized when AI is used as targeted scaffolding within sound pedagogy, transparent governance, and subgroup-sensitive evaluation.}
\label{fig:benefits_risks}
\end{figure}
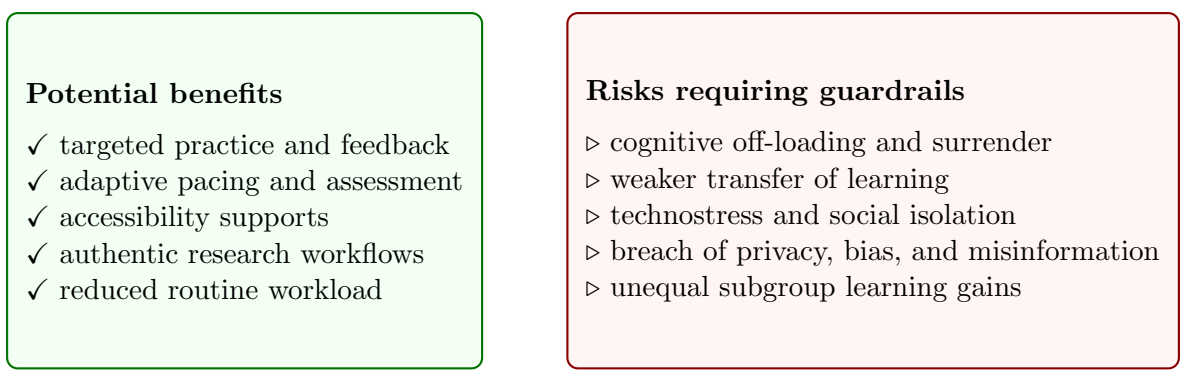

Figure~\ref{fig:benefits_risks} summarizes the central balance: AI can expand feedback, access, and research participation, but only if it is implemented with guardrails that preserve human reasoning, privacy, social learning, and equity in outcomes.

\section{Challenges and Risks}

AI's benefits come with significant risks that must be addressed through careful instructional design, governance, and equitable implementation. The most important risks are not only technical; they are cognitive, social, ethical, and institutional.

\subsection{Cognitive risks: off-loading, surrender, and weak transfer}

Researchers have documented a transfer-of-learning problem: students who rely heavily on AI assistance may perform well on the practiced task yet struggle to apply knowledge independently in new contexts~\cite{Bastani2024,Fan2024,Larson2024}. This distinction is important because the phrase \emph{transfer learning} also has a technical meaning in ML; here we use \emph{transfer of learning} to mean students' ability to apply knowledge and reasoning to unfamiliar problems.

A related risk is cognitive off-loading: strategically shifting mental work to an external tool. Off-loading can be beneficial when it frees cognitive resources for higher-level reasoning, but it becomes harmful when students stop practicing the reasoning they are supposed to learn. Shaw and Nave describe a deeper risk to learning, \emph{cognitive surrender}: accepting AI-generated outputs with limited scrutiny, allowing external artificial cognition---what they call ``System~3''---to bypass or suppress learners' own intuitive and deliberative reasoning~\cite{ShawNave2026}. Their experimental evidence shows that AI assistance can increase accuracy when AI is correct, reduce accuracy when AI is faulty, and inflate confidence even after errors~\cite{ShawNave2026}. In educational settings, this implies that students may feel confident while developing incomplete understanding.

These risks make independent verification and reflective explanation central design requirements. AI tutors and generative tools should ask students to explain assumptions, compare solutions, check outputs, and solve related problems without assistance. They should also be evaluated for sustained transfer, not only immediate task performance.

\subsection{Social, emotional, and self-regulation risks}

Over-reliance on AI tutors may contribute to digital fatigue, loneliness, and reduced face-to-face interaction, potentially weakening students' interpersonal skills, emotional intelligence, and sense of social support~\cite{Lai2023,Fan2024}. Increased screen time and continuous dependence on AI-mediated support may also contribute to \emph{technostress}: stress caused by persistent demands from digital technologies~\cite{Lai2023,Fan2024}. Students may also struggle to develop self-regulation---planning, monitoring, and adjusting one's own learning---if AI tools provide constant direction without requiring independent goal setting or reflection.

\subsection{Ethical, privacy, and governance risks}

Generative AI systems raise data privacy concerns, algorithmic bias, misinformation risks, and transparency challenges~\cite{UNESCOGenAI2023,OECD2025}. Student data can be misused if safeguards are weak, and opaque systems can erode trust when students and educators cannot understand why feedback or recommendations were produced. Ethical governance therefore requires clear policies on data collection, consent, transparency, accountability, human review, and appropriate use of AI-generated content.

\subsection{Equity risks beyond access}

Equity risks extend beyond the digital divide. Even when all students have access to the same AI platform, the platform may produce unequal benefits if it is better aligned with the language practices, prior knowledge, cultural contexts, or curricular resources of some groups than others. Kaufman et al. report uneven adoption of AI tools among U.S. teachers and principals~\cite{Kaufman2025}, and the Stanford Center for Racial Justice warns that AI could widen racial disparities if not addressed through inclusive design and equitable resource allocation~\cite{Pham2024}. For this reason, AI adoption should be evaluated with disaggregated evidence on learning gains, transfer performance, persistence, and confidence calibration, not only with participation counts or usage logs. Subgroup analyses should be conducted only when data can be collected ethically, sample sizes are sufficient for responsible interpretation, and privacy protections are in place.

\subsection{Materials-informatics-specific risks}

In materials discovery education, AI can create scientific reasoning risks if students treat model outputs as discoveries rather than hypotheses requiring validation. Poor data provenance can hide experimental incompatibilities. Descriptor choices can encode nonphysical correlations. Random data splits can create leakage and inflated performance. Uncalibrated uncertainty can lead students to over-prioritize candidates outside the training distribution. These risks are not reasons to avoid AI; they are reasons to teach AI as part of a scientific workflow grounded in verification, uncertainty, and experimental feedback.

Table~\ref{tab:positive_negative} summarizes reported positive and negative effects of learning with LLM-based cognitive tutors.

\begin{table}[H]
\centering
\caption{Reported positive and negative effects of learning with large language model (LLM)-based cognitive tutors. Evidence varies by study design, task format, guardrails, and outcome measure; therefore, both average effects and subgroup effects should be evaluated.}
\label{tab:positive_negative}
\begin{tabularx}{\textwidth}{>{\raggedright\arraybackslash}X >{\raggedright\arraybackslash}X}
\toprule
\textbf{Potential positive effects} & \textbf{Potential negative effects} \\
\midrule
Improved academic achievement and learning gains, especially when feedback is structured and aligned with instruction~\cite{Deng2024,VanLehn2011,KulikFletcher2016}. & Inflated beliefs about understanding, overconfidence, and reduced critical reasoning when students accept outputs without verification~\cite{Fan2024,Kosmyna2025,ShawNave2026}. \\
Higher-order thinking may be supported when AI prompts explanation, comparison, and reflection~\cite{ChiWylie2014}. & Heavy reliance can weaken transfer of learning and reduce independent problem solving~\cite{Bastani2024,Lehmann2024,Bastani2025,Li2025}. \\
Personalized feedback and adaptive practice can increase engagement, persistence, and timely correction. & Continuous AI guidance can contribute to cognitive off-loading, cognitive surrender, technostress, and reduced self-regulation. \\
AI can expand access to practice and feedback where tutoring or advanced coursework is limited. & Access remains uneven, and learning gains may differ across schools, socio-economic groups, language backgrounds, disability status, and other student subgroups~\cite{Kaufman2025,Pham2024}. \\
\bottomrule
\end{tabularx}
\end{table}

\section{Strategies and Recommendations}

To harness AI's potential while mitigating risks, institutions should treat AI education as a coordinated instructional, technical, ethical, and evaluation challenge. Tables~\ref{tab:pedagogy_recommendations} and~\ref{tab:governance_recommendations} convert the main recommendations into implementation actions and measurable indicators.

\begin{table}[H]
\centering
\caption{Curriculum and pedagogy recommendations for AI education in support of AI-powered materials discovery.}
\label{tab:pedagogy_recommendations}
\small
\begin{tabularx}{\textwidth}{>{\raggedright\arraybackslash}p{0.24\textwidth} >{\raggedright\arraybackslash}X >{\raggedright\arraybackslash}p{0.27\textwidth}}
\toprule
\textbf{Recommendation} & \textbf{Implementation action} & \textbf{Evidence or metric} \\
\midrule
Adopt comprehensive AI literacy frameworks & Map program outcomes to technical, data, evaluation, and socio-technical competencies; align with UNESCO, OECD, and higher-education frameworks~\cite{Fengchun2024,OECD2025,Kassorla2024}. & Curriculum map; competency rubrics; pre/post AI-literacy assessment. \\
Implement a dual-track curriculum & Combine foundational AI courses with AI integration in disciplinary courses, including MI, chemistry, physics, engineering, and ethics. & Number of courses with AI outcomes; student performance on domain-specific AI tasks. \\
Sustain professional development & Build year-long professional learning communities, AI Book Clubs, and workshop-to-course translation teams~\cite{Lee2022,Zhang2023,Zhang2024}. & Instructor participation; revised assignments; confidence and practice-change surveys. \\
Apply constrained agency & Design AI tools to provide just-in-time scaffolding without taking over reasoning; require explanation, verification, and independent attempts~\cite{Bastani2025,Li2025}. & Transfer-task performance; reduced over-reliance; quality of student explanations. \\
\bottomrule
\end{tabularx}
\end{table}

\begin{table}[H]
\centering
\caption{Governance, equity, and evaluation recommendations for AI education in support of AI-powered materials discovery.}
\label{tab:governance_recommendations}
\small
\begin{tabularx}{\textwidth}{>{\raggedright\arraybackslash}p{0.24\textwidth} >{\raggedright\arraybackslash}X >{\raggedright\arraybackslash}p{0.27\textwidth}}
\toprule
\textbf{Recommendation} & \textbf{Implementation action} & \textbf{Evidence or metric} \\
\midrule
Protect social and collaborative learning & Use AI to support, not replace, peer interaction, project-based learning, mentoring, and expert feedback. & Peer collaboration measures; student belonging; group project outcomes. \\
Teach critical thinking and ethical design & Include bias, fairness, privacy, prompt engineering, model documentation, and civic implications of AI~\cite{Maloy2025,Veldhuis2025}. & Ethics case analyses; model cards; data sheets; critique of AI outputs. \\
Ensure governance and equity in outcomes & Establish policies for data privacy, transparency, accessibility, and disaggregated evaluation of learning gains and transfer~\cite{UNESCOGenAI2023,Pham2024}. & Access metrics plus subgroup learning gains, persistence, and confidence calibration. \\
Support interdisciplinary research and evaluation & Connect education researchers, cognitive scientists, ethicists, domain scientists, and technologists in continuous improvement cycles. & Published instruments; shared datasets; cross-institution comparisons; iterative redesign evidence. \\
\bottomrule
\end{tabularx}
\end{table}

Several principles cut across these recommendations. First, AI should be designed for \emph{constrained agency}: it should support learners without taking over the reasoning process. In practice, this means limiting AI interventions to just-in-time scaffolding, asking students to attempt reasoning before receiving answers, and requiring explanation and verification. Second, governance should be treated as part of pedagogy rather than as a separate compliance layer. Students should learn why privacy, documentation, transparency, and accountability matter for scientific work. Third, equity should be evaluated by outcomes that matter: conceptual understanding, transfer of learning, self-regulation, confidence calibration, persistence, and research readiness.

\begin{figure}[H]
\centering
\begin{tikzpicture}[
  font=\small,
  node distance=1.05cm and 1.05cm,
  center/.style={draw=blue!65!black, rounded corners, very thick, fill=blue!10, align=center, minimum width=3.6cm, minimum height=1.05cm},
  item/.style={draw=#1!65!black, rounded corners, thick, fill=#1!9, align=center, minimum width=3.05cm, minimum height=0.85cm},
  arrow/.style={-{Latex[length=2.0mm]}, thick, gray!75}
]
\node[center] (c) at (0,0) {\textbf{Responsible AI}\\\textbf{education for MI}};
\node[item=blue]   (lit)    at (-4.0, 1.7) {AI literacy\\frameworks};
\node[item=orange] (curr)   at (0, 2.25) {dual-track\\curriculum};
\node[item=green]  (pd)     at (4.0, 1.7) {professional\\development};
\node[item=purple] (agency) at (-4.0,-1.7) {constrained\\agency};
\node[item=cyan]   (social) at (0,-2.25) {social and\\collaborative learning};
\node[item=red]    (eval)   at (4.0,-1.7) {interdisciplinary\\evaluation};
\begin{scope}[on background layer]
  \node[draw=gray!65, rounded corners, dashed, thick, fill=gray!6, inner sep=10pt,
        fit=(lit)(curr)(pd)(agency)(social)(eval)] (gov) {};
  \node[fill=white, text=gray!70!black] at (gov.north) {\textbf{governance + outcome equity layer}};
\end{scope}
\foreach \n in {lit,curr,pd,agency,social,eval}{\draw[arrow] (\n) -- (c); \draw[arrow] (c) -- (\n);}
\end{tikzpicture}
\caption{Integrated strategy set for AI education in materials informatics. Governance and outcome equity surround the full system because privacy, transparency, accessibility, subgroup-sensitive evaluation, and accountability apply to every curricular and pedagogical component.}
\label{fig:strategies}
\end{figure}
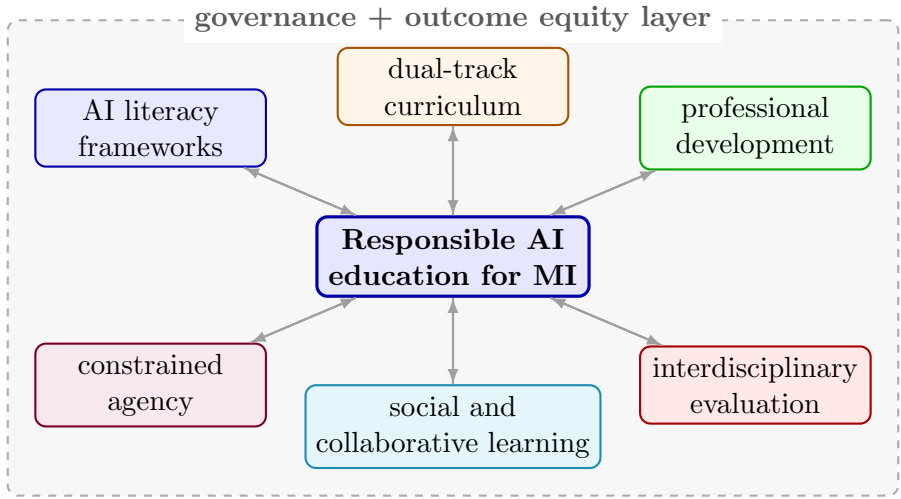

Figure~\ref{fig:strategies} provides a compact framework for implementation and shows governance plus outcome equity as a system-level layer rather than a single isolated task. The key point is that no single intervention is sufficient. A course module without faculty development will not scale; a platform without governance may create privacy and equity risks; access without learning gains will not close gaps; and AI literacy without authentic scientific tasks will not prepare students for AI-powered materials discovery.

\section{Concrete Curriculum Artifacts for Materials Informatics Education}
\label{sec:curriculum_artifacts}

To operationalize the dual-track model and the MI competency needs described above, this section provides reusable curriculum artifacts: an example eight-week module sequence, three representative assignments, and a compact assessment rubric. These artifacts are designed to be deployable within existing materials science curricula as an upper-division elective module, graduate mini-course, workshop, bootcamp sequence, or cross-institution short course. They can be adapted to local constraints such as credit hours, computing access, student preparation, and available datasets.

\subsection{Example eight-week MI module sequence}

Table~\ref{tab:module_sequence} outlines an eight-week sequence aligned to the AI-powered materials discovery workflow: data, representation, modeling, validation, decision making, and experimental feedback.

\begin{table}[H]
\centering
\caption{Example eight-week module sequence for MI education. Each week includes a primary topic and measurable learning objectives.}
\label{tab:module_sequence}
\small
\begin{tabularx}{\textwidth}{>{\centering\arraybackslash}p{0.07\textwidth} >{\raggedright\arraybackslash}p{0.28\textwidth} >{\raggedright\arraybackslash}X}
\toprule
\textbf{Week} & \textbf{Topic} & \textbf{Learning objectives: students can \ldots} \\
\midrule
1 & MI workflow and data foundations & describe the closed-loop MI workflow; identify experiment, simulation, and literature data; assess missingness, units, provenance, and metadata quality. \\
2 & Materials representations and featurization & construct and justify composition- and structure-based descriptors; explain expressivity vs. data-demand tradeoffs; document descriptor choices. \\
3 & Supervised ML basics & implement baseline regression or classification models; apply train/validation/test splits; compute and interpret MAE, RMSE, $R^2$, F1 score, or AUC with confidence intervals. \\
4 & Generalization, leakage, and dataset shift & diagnose leakage; compare random, grouped, and time-aware splits; test robustness under composition-space or process-space shifts. \\
5 & Uncertainty quantification and calibration & estimate predictive uncertainty; evaluate calibration; communicate uncertainty for materials decisions. \\
6 & Physics-informed and scientific AI & incorporate physical constraints or priors; detect implausible predictions; compare purely statistical and physics-informed models. \\
7 & Active learning and closed-loop design & design an acquisition function; simulate an active-learning loop; analyze sample efficiency under small-data constraints. \\
8 & Reproducibility, ethics, and deployment & produce a model card or data sheet; discuss privacy, bias, access, and subgroup impacts; present an end-to-end MI mini-project. \\
\bottomrule
\end{tabularx}
\end{table}

\subsection{Example assignments}

The following assignments are feasible in Jupyter notebooks using public datasets such as Materials Project-derived datasets, OQMD subsets, or instructor-provided experimental datasets. They can be scaled for undergraduate or graduate levels. These assignments are AI-literacy-relevant because they require students to move beyond tool execution toward core AI-literacy practices: documenting data provenance, justifying representations, evaluating model validity, interpreting uncertainty, detecting failure modes, and connecting AI outputs to scientific judgment.

\paragraph{Assignment A: Descriptor benchmarking and justification.} Students compare at least three descriptor families, such as Magpie composition statistics, simple structural features, and learned embeddings, on a fixed target property. Deliverables include: (i) a rationale for descriptor choice, (ii) a performance comparison with uncertainty, and (iii) an error analysis identifying regimes where each descriptor fails. This assignment targets the MI bottleneck of user-dependent featurization.

\paragraph{Assignment B: Leakage and dataset-shift diagnosis.} Students intentionally construct a naive split that produces inflated performance, such as a random split with near-duplicate compositions, and then correct it using grouped splits by chemical system or another scientifically justified split. Deliverables include: (i) evidence of leakage, (ii) a revised evaluation protocol, and (iii) discussion of how conclusions change under realistic generalization goals.

\paragraph{Assignment C: Uncertainty-aware active learning loop.} Students train an ensemble model, compute uncertainty, and implement an acquisition rule to select the next experiments or simulations to run. Deliverables include: (i) a learning curve showing performance vs. number of labels, (ii) comparison to random sampling, and (iii) a one-page memo explaining how uncertainty would guide real laboratory decisions.

\subsection{Rubric and assessment instrument}

Table~\ref{tab:mi_rubric} provides a compact rubric for grading projects or for pre/post evaluation of MI competencies. Scores can be tracked longitudinally across courses or bootcamps to evaluate learning impact. The rubric aligns with the framework proposed in this paper by translating workflow-aligned AI literacy for materials discovery---data provenance, representation, validation, uncertainty, physics-informed reasoning, reproducibility, ethics, and equity---into observable performance indicators.

\begin{table}[H]
\centering
\caption{Compact rubric for assessing MI competencies. Each criterion can be scored on a 0--2 scale: 0=insufficient, 1=adequate, 2=strong.}
\label{tab:mi_rubric}
\small
\begin{tabularx}{\textwidth}{>{\raggedright\arraybackslash}p{0.31\textwidth} >{\raggedright\arraybackslash}X}
\toprule
\textbf{Competency} & \textbf{Performance indicators} \\
\midrule
Data quality and provenance & Units consistent; missingness handled; sources documented; metadata recorded; train/test separation justified. \\
Representation and featurization & Descriptors selected with domain rationale; ablation or benchmarking performed; limitations and invariances stated. \\
Modeling and validation & Appropriate baseline; hyperparameters documented; metrics correct; confidence intervals or uncertainty reported; leakage avoided. \\
Robustness and shift awareness & Evaluation includes realistic splits; failure modes under shift identified; mitigation proposed. \\
Physics-informed reasoning & Physical constraints considered; implausible predictions flagged; interpretation connects to materials principles. \\
Reproducibility and communication & Code runs end-to-end; results reproducible; model card or data sheet included; uncertainty communicated for decisions. \\
Ethics and equity awareness & Privacy and governance considered; access constraints acknowledged; subgroup impacts and differential learning gains considered; responsible-use recommendations stated. \\
\bottomrule
\end{tabularx}
\end{table}

\subsection{Assessment and evaluation plan}
\label{sec:assessment_evaluation}

For publication-quality evidence and continuous improvement, MI education initiatives should be paired with an explicit evaluation plan that measures (i) learning gains, (ii) applied competence in authentic MI tasks, (iii) longer-term outcomes, and (iv) equity of access. Because many MI programs are introduced as modules, bootcamps, or course inserts rather than full degree redesigns, the evaluation approach should be lightweight, repeatable, and aligned to the closed-loop materials discovery workflow.

\paragraph{Pre/post instruments adapted to materials informatics.} Short pre/post assessments should target MI-specific AI literacy rather than generic coding familiarity. Instrument domains should include data provenance, representation choices, model validation, uncertainty and calibration, and physical plausibility. Items can mix multiple-choice, short-answer explanations, and ``debug-the-claim'' prompts in which students inspect a model claim and justify whether it is credible given the evaluation protocol.

\paragraph{Performance tasks for authentic MI competencies.} Evaluation should include tasks that mirror real MI practice: diagnosing data leakage or dataset shift; comparing descriptor families for the same target property; and translating uncertainty into experimental prioritization. Rubrics such as Table~\ref{tab:mi_rubric} can support consistent scoring across courses and institutions.

\paragraph{Longitudinal indicators and research readiness.} Programs should track longer-term indicators when feasible, including retention in MI-related pathways, participation in research projects, posters or publications, open-source contributions, and reproducibility behaviors such as version control, documented datasets, reusable notebooks, and transparent reporting. In research settings, a practical indicator is whether students can independently reproduce a published MI baseline and then extend it with a justified improvement.

\paragraph{Equity metrics: access and differential outcomes.} Evaluation should measure whether AI/MI education expands opportunity or reinforces disparities. Recommended metrics include access to computing resources and datasets, participation rates by subgroup, differential learning gains between subgroups, transfer to unfamiliar tasks, persistence in MI pathways, and confidence calibration across demographics, institutions, rurality, and prior-preparation levels where such data can be collected ethically and appropriately. Subgroup analyses should be conducted only when sample sizes are sufficient for responsible interpretation, privacy protections are in place, and the results will be used to improve support rather than stigmatize learners or institutions. The central equity question is not only whether underserved students can participate, but whether they benefit at least as much as better-resourced peers on conceptual understanding, independent problem solving, scientific reasoning, and research readiness.

\paragraph{Iterative improvement.} Evaluation results should be used formatively: to refine modules, adjust scaffolding, improve guardrails around AI tool use, and target instructor professional development. Reporting a minimal evaluation package---instruments, task prompts, rubric, and anonymized summary results---will improve reproducibility and allow cross-institution comparison of MI education approaches.

\section{Limitations and Scope}

This paper synthesizes existing research and workshop-informed practice; it does not present new empirical data from an implemented curriculum. Some cited sources are peer-reviewed studies or meta-analyses, while others are surveys, policy documents, workshop contributions, or professional guidance. We have therefore treated them according to their evidentiary role: surveys describe adoption and perception, policy documents guide governance, and experimental studies or syntheses support stronger claims about learning effects. A second limitation is that AI tools and institutional policies are evolving rapidly, so specific technologies and adoption rates will change. The framework proposed here is intended to be durable because it focuses on underlying competencies---data provenance, representation, validation, uncertainty, scientific reasoning, governance, and subgroup-sensitive evaluation---rather than on any single AI platform. A third limitation is that subgroup-sensitive evaluation is essential but methodologically demanding: small samples, missing demographic data, and privacy constraints can limit what can be responsibly inferred. Finally, implementation will vary by institutional resources, student preparation, faculty expertise, and local data infrastructure. The curriculum artifacts should therefore be viewed as adaptable templates rather than fixed prescriptions.

\section{Conclusion}

AI presents transformative possibilities for education, including personalized feedback, expanded access, authentic research participation, and stronger preparation for AI-powered materials discovery. Yet these benefits are accompanied by risks related to cognitive off-loading, cognitive surrender, weak transfer of learning, privacy, bias, social isolation, and inequitable subgroup outcomes. A balanced approach---grounded in robust AI literacy, effective pedagogy, ethical design, transparent governance, and careful evaluation---is essential.

The central message of this paper is that AI education for materials informatics should not remain generic or tool-centered; it should be embedded in domain practice and should build scientific judgment. Students should learn not only how to use AI, but how to question it: where the data came from, what assumptions are embedded in a representation, whether validation is credible, what uncertainty remains, how predictions connect to physical mechanisms, and whether AI-supported learning benefits all students. By investing in teacher preparation, dual-track curricula, constrained agency, social learning, responsible governance, and outcome-oriented equity, institutions can ensure that AI becomes a force for educational improvement rather than another source of inequity. For materials discovery, this means preparing students to become critical thinkers, ethical designers, and innovative scientists who can use AI productively while preserving human reasoning, experimental judgment, and responsibility.

\section*{Acknowledgments}
We acknowledge support from the National Science Foundation under Grants
\textcolor{blue}{No. OIA-2427805}, OISE-1743790, PHYS-2117774,  PHYS-2310027, OIA-2437416, and OIA-2412055, and from the U.S.\ Department of Energy under Grants
No.~DE-SC0024519 and DE-SC0004768, as well as support from a research center
funded by the State of South Dakota.



    \bibliographystyle{unsrt}
    \bibliography{ai_education_white_paper}

\end{document}